\documentclass[amsmath,amssymb,preprint,superscriptaddress]{revtex4-1}

\usepackage[dvips]{graphicx} % for figures
\usepackage{amsfonts}
\usepackage{amssymb}
\usepackage{amscd}
\usepackage{amsmath}    % need for subequations
\usepackage{enumerate}
\usepackage{epsfig}
\usepackage{subfigure}
\usepackage{xcolor}
\usepackage{array}
\usepackage{amsthm}

\begin{document}

\title{Time-Frequency Transfer through a 70 dB Free Space Channel: Towards Satellite-Ground Time Dissemination}

\author{Qi Shen\textsuperscript{*}}
\author{Jian-Yu Guan\textsuperscript{*}}
\author{Ting Zeng}
\author{Qi-Ming Lu}
\author{Liang Huang}
\author{Yuan Cao}
\author{Jiu-Peng Chen}
\affiliation{Shanghai Branch, National Laboratory for Physical Sciences at Microscale and Department of Modern Physics, University of Science and Technology of China, Shanghai 201315, China}
\affiliation{CAS Center for Excellence and Synergetic Innovation Center in Quantum Information and Quantum Physics, University of Science and Technology of China, Shanghai, 201315, China}
\author{Tian-Qi Tao}
\author{Jin-Cai Wu}
\affiliation{CAS Center for Excellence and Synergetic Innovation Center in Quantum Information and Quantum Physics, University of Science and Technology of China, Shanghai, 201315, China}
\affiliation{Laboratory of Space Active Opto-Electronic Technology, Shanghai Institute of Technical Physics, Chinese Academy of Sciences, Shanghai 200083, China}
\author{Lei Hou}
\affiliation{Key Laboratory of Time and Frequency Primary Standards, National Time Service Center, Chinese Academy of Sciences, Xi’an 710600, China}
\author{Sheng-Kai Liao}
\author{Ji-Gang Ren}
\author{Juan Yin}
\affiliation{Shanghai Branch, National Laboratory for Physical Sciences at Microscale and Department of Modern Physics, University of Science and Technology of China, Shanghai 201315, China}
\affiliation{CAS Center for Excellence and Synergetic Innovation Center in Quantum Information and Quantum Physics, University of Science and Technology of China, Shanghai, 201315, China}
\author{Jian-Jun Jia}
\affiliation{CAS Center for Excellence and Synergetic Innovation Center in Quantum Information and Quantum Physics, University of Science and Technology of China, Shanghai, 201315, China}
\affiliation{Laboratory of Space Active Opto-Electronic Technology, Shanghai Institute of Technical Physics, Chinese Academy of Sciences, Shanghai 200083, China}
\author{Hai-Feng Jiang}
\affiliation{Shanghai Branch, National Laboratory for Physical Sciences at Microscale and Department of Modern Physics, University of Science and Technology of China, Shanghai 201315, China}
\affiliation{CAS Center for Excellence and Synergetic Innovation Center in Quantum Information and Quantum Physics, University of Science and Technology of China, Shanghai, 201315, China}
\affiliation{Key Laboratory of Time and Frequency Primary Standards, National Time Service Center, Chinese Academy of Sciences, Xi’an 710600, China}
\author{Cheng-Zhi Peng}
\author{Qiang Zhang}
\author{Jian-Wei Pan}
\affiliation{Shanghai Branch, National Laboratory for Physical Sciences at Microscale and Department of Modern Physics, University of Science and Technology of China, Shanghai 201315, China}
\affiliation{CAS Center for Excellence and Synergetic Innovation Center in Quantum Information and Quantum Physics, University of Science and Technology of China, Shanghai, 201315, China}

%\author{Qi Shen\textsuperscript{1,2,*}, Jian-Yu Guan\textsuperscript{1,2,*}, Ting Zeng\textsuperscript{1,2}, Qi-Ming Lu\textsuperscript{1,2}, Liang Huang\textsuperscript{1,2}, Yuan Cao\textsuperscript{1,2}, Jiu-Peng Chen\textsuperscript{1,2}, Tian-Qi Tao\textsuperscript{2,3}, Jin-Cai Wu\textsuperscript{2,3}, Lei Hou\textsuperscript{4}, Sheng-Kai Liao\textsuperscript{1,2}, Ji-Gang Ren\textsuperscript{1,2}, Juan Yin\textsuperscript{1,2}, Jian-Jun Jia\textsuperscript{2,3}, Hai-Feng Jiang\textsuperscript{1,2,4}, Cheng-Zhi Peng\textsuperscript{1,2}, Qiang Zhang\textsuperscript{1,2}, Jian-Wei Pan\textsuperscript{1,2}}

%\begin{affiliations}
%    \item Shanghai Branch, National Laboratory for Physical Sciences at Microscale and Department of Modern Physics, University of Science and Technology of China, Shanghai 201315, China
%    \item CAS Center for Excellence and Synergetic Innovation Center in Quantum Information and Quantum Physics, University of Science and Technology of China, Shanghai, 201315, China
%    \item Key Laboratory of Space Active Opto-Electronic Technology, Shanghai Institute of Technical Physics, Chinese Academy of Sciences, Shanghai 200083, China
%    \item Key Laboratory of Time and Frequency Primary Standards, National Time Service Center, Chinese Academy of Sciences, Xi’an 710600, China
%    \\
%    $^{*}$These authors contributed equally to this work.
%\end{affiliations}
\baselineskip24pt

\begin{abstract}
Time and frequency transfer lies at the heart of the field of metrology. Compared to current microwave dissemination such as GPS~\cite{el2002introduction}, optical domain dissemination can provide more than one order of magnitude in terms of higher accuracy~\cite{droste2013optical,chiodo2015cascaded}, which allows for many applications such as the redefinition of the second, tests of general relativity and fundamental quantum physics, precision navigation and quantum communication~\cite{Riehle_2018,riehle2015towards,kolkowitz2016gravitational,derevianko2014hunting,liu2019experimental}. Although optical frequency transfer has been demonstrated over thousand kilometers fiber lines~\cite{droste2013optical}, intercontinental time comparison and synchronization still requires satellite free space optical time and frequency transfer. Quite a few pioneering free space optical time and frequency experiments have been implemented at the distance of tens kilometers~\cite{deschenes2016synchronization,sinclair2016synchronization} at ground level. However, there exists no detailed analysis or ground test to prove the feasibility of satellite-based optical time-frequency transfer. Here, we analyze the possibility of this system and then provide the first-step ground test with high channel loss. We demonstrate the optical frequency transfer with an instability of $10^{-18}$ level in 8,000 seconds across a 16-km free space channel with a loss of up to 70~dB, which is comparable with the loss of a satellite-ground link at medium earth orbit (MEO) and geostationary earth orbit (GEO).
\end{abstract}

\maketitle

\section{Main}
%State-of-art cold atom optical clocks can reach $10^{-19}$ in 10,000 seconds, which is 2-3 orders of magnitude performance over Cesium microwave clocks~\cite{liu2016improvement} and thus are the best candidates for the next generation of the definition of time. Meanwhile, widely used satellite based microwave dissemination system, like GPS, Gallileo or Beidou can only distribute time and frequency signal with a stability of $10^{-16}$ in 10,000 seconds, not compatible with the optical clock. Naturally, a global optical time and frequency dissemination network is required to synchronize and compare all the cold-atom optical clocks, which can find immediate applications in redefinition of seconds, test of new physics and quantum metrology.

Fiber-based optical time-frequency dissemination networks have been demonstrated as a solution for comparisons between fixed sites connected by a fiber link~\cite{ma1994delivering,droste2013optical,chiodo2015cascaded}. However, since it is difficult to apply fiber links in locations such as mountainous\cite{katori2011optical} and marine regions, while the limitations for intercontinental dissemination are obvious, satellite-based optical time-frequency dissemination is a reasonable option. Traditional satellite-based links exhibit an optimum frequency instability of around $1\times10^{-15}$ for a day, which is limited according to the resolution of the microwave carrier~\cite{bauch2005comparison}. To improve this resolution, time transfer by laser link (T2L2) with short laser pulses carrier has been explored. However, the limitation due to the response speed of photoelectric detections makes it difficult to reach a level of $10^{-17}$ in terms of stability\cite{Michalek2015, Samain2015}. In fact, this performance is comparable to that of the new generation microwave links with multi-carriers for the ACES project\cite{Exertier2019, Cacciapuoti2017}. The planned European Laser Timing (ELT+) optical link for the I-SOC project is aimed at providing a frequency transfer uncertainty of below orders of $10^{-18}$ for integration times longer than 10 days\cite{Cacciapuoti2017}. At present, no satellite link can transfer time-frequency signals generated by the best optical frequency standards with a stability of below $1\times10^{-16}/\sqrt{\tau}$, where $\tau$ is the averaging time in seconds\cite{Oelker2019}.
%Compared to the above techniques, the recently developed technique of combining frequency comb and linear optical sampling (LOS)\cite{coddington2009rapid} can reach orders of $10^{-18}$ within 10,000 seconds in ground with a static\cite{deschenes2016synchronization,sinclair2016synchronization} or movable receiver \cite{bergeron2019femtosecond}.

\begin{figure}
\centering
\includegraphics[width=0.8\columnwidth]{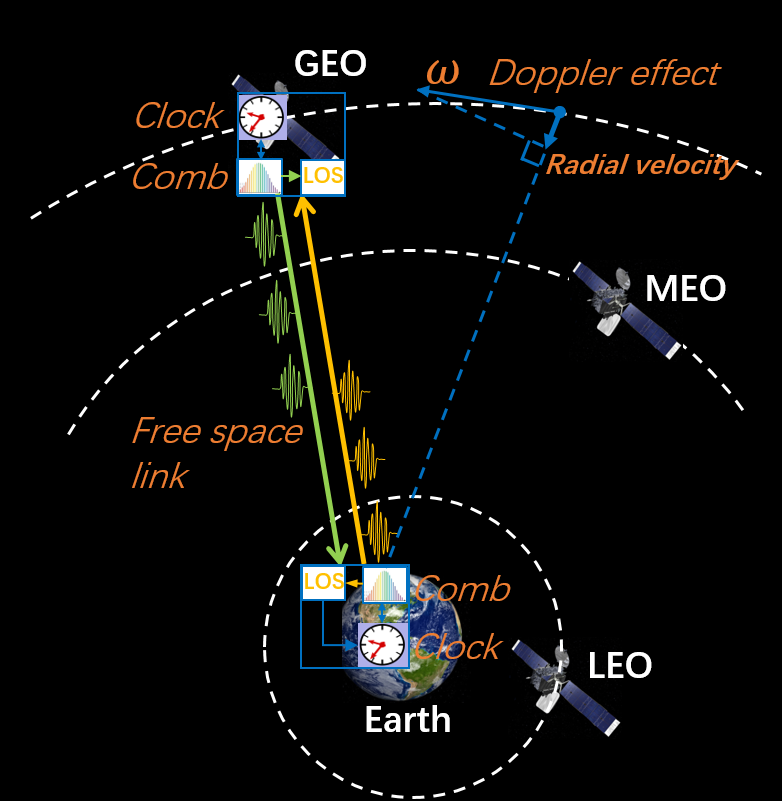}
\caption{Satellite-ground optical time-frequency dissemination diagram with different orbit types.}
\label{Fig:Sketch}
\end{figure}

 A pioneering technique that combines frequency combs and linear optical sampling (LOS)\cite{coddington2009rapid} based on optical interference, which avoids any photoelectronic response limitation, can reach orders of $10^{-18}$ within 1,000 seconds in grounds with a static\cite{deschenes2016synchronization,sinclair2016synchronization} or a movable platform \cite{bergeron2019femtosecond}. Here, we first analyze this new route's feasibility for satellite-based optical frequency dissemination. Fig.~\ref{Fig:Sketch} shows a simplified version of the satellite-ground optical time-frequency dissemination diagram. Two frequency combs were installed, one on the satellite and another on the ground station, with the two combs connected by a satellite-ground free space link consisting of two telescopes. Both sites could send and receive comb light through the telescopes, while at each site, the received light could be coupled into single-mode fiber to beat with the local comb. Then, the difference of the beat signals from the two sites could be calculated and used as the feedback signal to synchronize the optical frequencies~\cite{deschenes2016synchronization}.

\begin{figure}
\centering
\includegraphics[width=0.9\columnwidth]{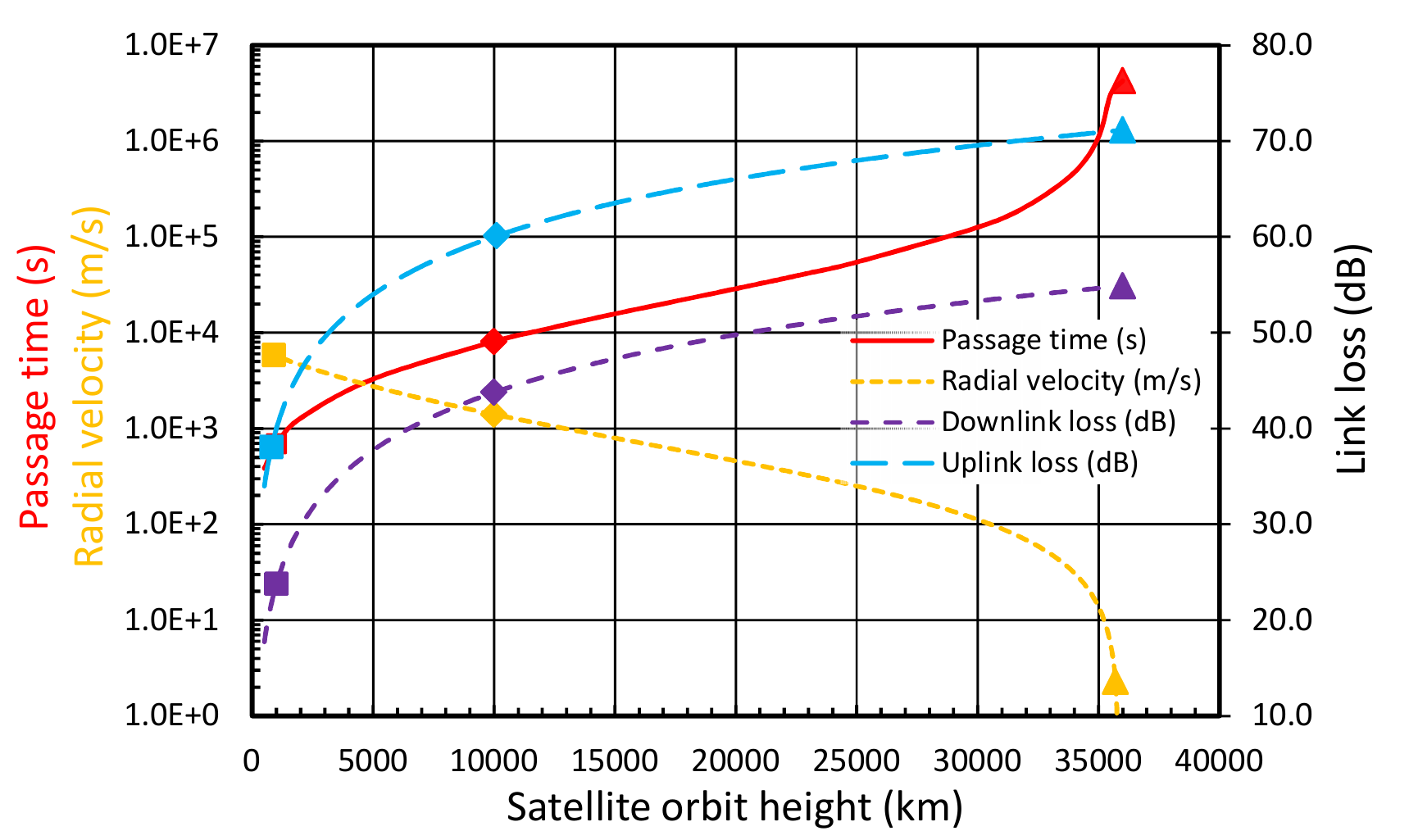}
\caption{Satellite-ground link loss, radial velocity, and passage time with satellite orbit height. The satellite orbit is assumed to be circular with a zero-degree inclination. The ground station is set in the equator line and the field of view is 165$^{\circ}$ from the minimum elevation angle of 15$^{\circ}$. The radial velocity is calculated with the satellite at the elevation angle of 15$^{\circ}$ from the horizon plane of the ground station. The points with solid squares, rhombuses and triangles labels represent typical LEO, MEO and GEO satellites, respectively.}
\label{Fig:height}
\end{figure}

Transmission loss and the Doppler shift are two key factors for consideration. The transmission loss consists of both the free space link loss and the local optic loss. A good optical frequency transfer will involve low loss and a small Doppler shift. However, there is a tradeoff between these two factors. When the satellite orbit height is low, the link loss is also low. However, the speed of the satellite is relatively high, which results in a large Doppler shift. With an increase in orbit height, the link loss becomes larger, while its speed decreases and the Doppler shift becomes smaller. When the height reaches the geostationary earth orbit (GEO), the satellite is relatively still in relation to the earth and the Doppler shift almost disappears. Fig.~\ref{Fig:height} shows how the link loss and Doppler shift values vary according to satellite orbit height, as well as the passage time. Table 1 provides the typical values for different orbit types (see Methods).

For a low earth orbit (LEO), the typical height is around 1,000~km and the maximum radial velocity is around $5.6~km/s$. The Doppler shift for the optical frequency of 200 THz will reach up to 3.73~GHz. For a comb with a repetition rate of 200~MHz, the received comb will move across multiple comb spans. This leads to more difficulties for the continuous working of the LOS. Meanwhile, a larger Doppler effect leads to a larger Delay-Doppler effect. However, there currently exists no experimental validation method for such a large Doppler effect. Moreover, compared with LEO, the MEO and GEO have much longer passage times, which allows for more time for accumulating data and reducing the instability.

% For a low earth orbit (LEO), the typical height is around 500 km and thus the upper bound for the feedback bandwidth is 300 kHz, which is set by the speed of the light in vacuum. We can tell from the table I, this bandwidth is unfortunately lower than the doppler shit, which turns out to be difficult for compensation. For MEO or GEO, when the orbit is higher than xx km, the doppler shift will be lower than the feedback bandwidth. Moreover, comparing to LEO, the medium earth orbit (MEO)or GEO orbit also has much longer passage time, which will bring more time for accumulating data and reducing the instability.

In view of this, the MEO and GEO should be suitable options for the target orbit. For these orbits, the biggest obstacle is the high link loss between satellite and ground. As is clear from Fig.~\ref{Fig:height} and Table 1, the link efficiency is quite different between the downlink and the uplink, which represent satellite-to-ground and ground-to-satellite, respectively. In the two-way satellite-ground time dissemination, both links are employed simultaneously. However, the optical uplink loss is much higher than the downlink loss, which is manily due to the different divergence angles and the different efficiency of single-mode fiber coupling. With the typical system parameters~\cite{ren2017ground, Bourgoin2013, jovanovic2016efficiently}, the maximum link attenuation is estimated to be as large as 71 dB for the uplink of a GEO with a 36,000 km satellite-ground distance, while it is around 60 dB for the uplink of the MEO with a 10,000 km distance (see Methods for detailed analysis of satellite-ground link loss).

\begin{table}\label{Table:Loss}
\centering
\begin{tabular}{|c|c|c|c|c|c|c|}
\hline
Satellite		& \multicolumn{2}{c|}{LEO} & \multicolumn{2}{c|}{MEO} 	& \multicolumn{2}{c|}{GEO} \\ \hline
Height(km)		& \multicolumn{2}{c|}{1,000} & \multicolumn{2}{c|}{10,000} 	& \multicolumn{2}{c|}{36,000} \\ \hline
Link				& Downlink & Uplink & Downlink & Uplink & Downlink & Uplink \\ \hline
Total loss (dB) & 23.8 & 40 & 43.8 & 60 & 54.9 & 71.1 \\ \hline
Radial velocity (km/s) & \multicolumn{2}{c|}{$\approx 6$} & \multicolumn{2}{c|}{$\approx 1$} & \multicolumn{2}{c|}{$\approx 0.001$} \\ \hline
Passage time (s) & \multicolumn{2}{c|}{$\approx 10^{3} $} & \multicolumn{2}{c|}{$\approx 10^{4}$} & \multicolumn{2}{c|}{$\approx 10^{6}$}  \\ \hline
\end{tabular}
\caption{The estimation of link loss, radial velocity, and passage time for each type of satellites. The loss becomes higher according to orbit altitude, while the radial velocity becomes smaller. The Doppler effect is proportional to the radial velocity. The divergence angle is 4 urad and 15 urad for downlink and uplink, respectively. The single-mode fiber coupling efficiency is 15\% and 5\% for downlink and uplink, respectively. The aperture of the receive telescope is 1 m and the telescope efficiency is 80\%, for both satellite and ground. }
\end{table}

To verify the feasibility of the frequency transmission for MEO and GEO satellites, we constructed a free-space channel consisting of a 16~km atmosphere and an extra adjustable attenuator. In our experiment, the link loss was set from 52 dB to 70~dB, with the loss capable of efficiently mimicking the situation of a satellite-ground link. While the main principle of free-space two-way time synchronization is the same as that for fibers circumstances, the difficulty is much greater with the former because of the high attenuation of the link and the random signal fades. In fact, the random signal fades meant that our measurement method had to have a large ambiguous length. The frequency transfer method using ultra-stable lasers is not suitable for long-term free-space frequency transfer due to its poor recover ability and the uncertainty channel. The femtosecond optical comb is thus a better candidate because it can provide a much larger ambiguity length that enables a good recover ability, and because it can achieve a better signal-to-noise ratio because of its much higher peak power. Previous works have demonstrated that the LOS using combs is a suitable tool for frequency transfer in free space~\cite{giorgetta2013optical,deschenes2016synchronization}, and a similar method was thus used in our experiment. The detailed principle and algorithm can be found in Methods part.

\begin{figure}
\centering
\includegraphics[width=0.9\columnwidth]{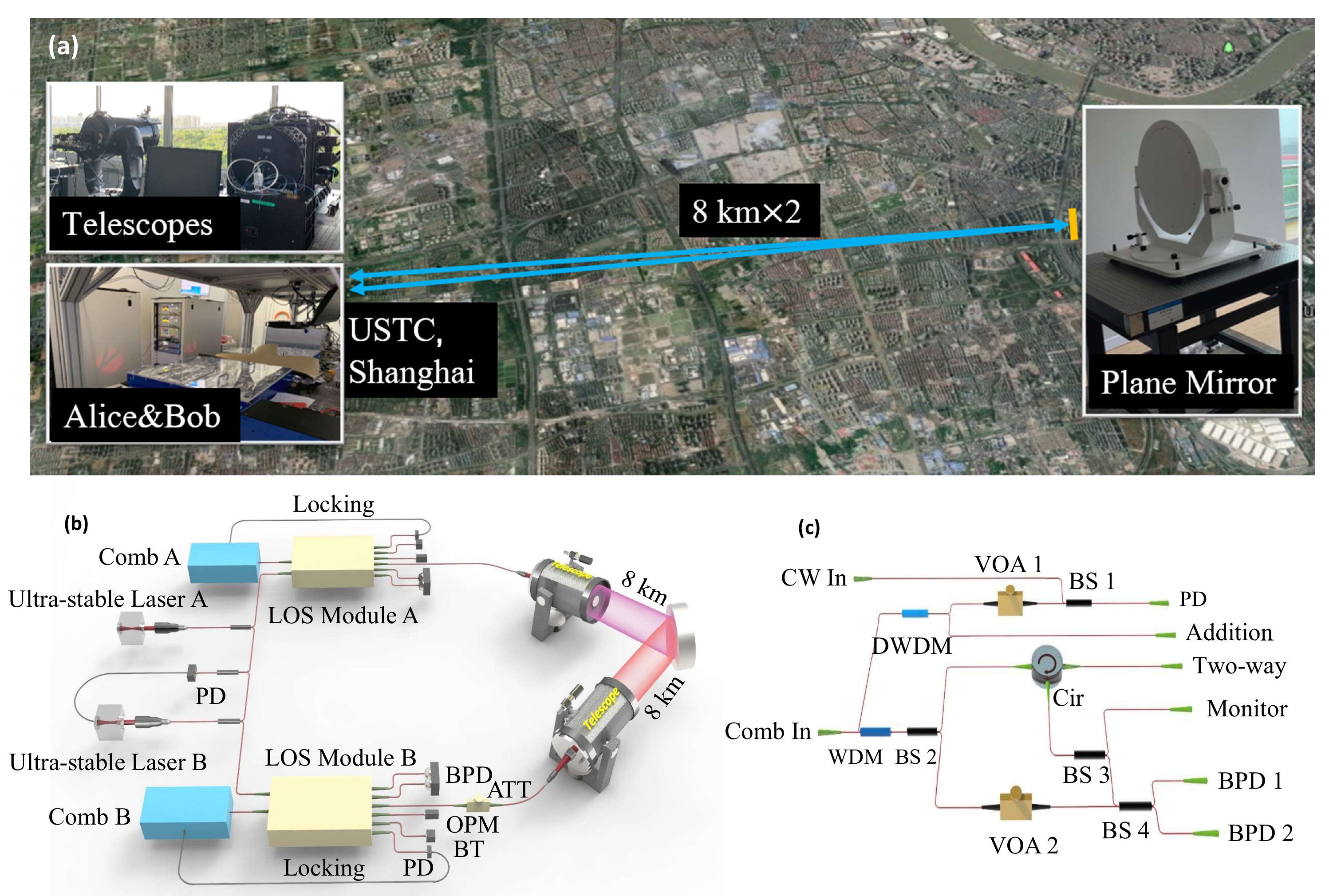}
\caption{ (a). Transfer takes place at Pudong, Shanghai between two co-located sites Alice and Bob with synchronized reference clock sources. The sites are linked by two 70 m optical fibre paths from the laboratory to each free-space launch for the 16 km air path; (b). Our experimental setup. PD: Photon diode. BT: Beam trap. ATT: Attenuator. OPM: Optical power meter. BPD: Balanced photon diode. LOS Module: a compact box for linear optical sampling; (c) the detailed structure in our LOS Module. WDM: A 20-nm band-pass filter centered at 1,520~nm. DWDM: A 0.8-nm narrow filter centered at 1,550.12~nm. Cir: Circulator. VOA 1 and VOA 2: Attenuator. BS 1: 50:50 beam splitter for beat between CW laser and comb. BS 2: 99:1 beam splitter, its 99\% port is connected to the circulator. BS 3: 95:5 beam splitter, its 5\% port is connected to the monitor port. BS 4: 50:50 beam splitter for beat between local comb and signal.}
\label{Fig:Setup}
\end{figure}

% experiment part
To effectively evaluate the stability of the time transfer, we used the same clock at two sides, which are called Alice and Bob, respectively. In our experiment, both Alice and Bob had an ultra-stable laser as the reference clock source. The two lasers were locked to each other in advance to ensure synchronization. The comb on each side was then locked to the corresponding ultra-stable laser, meaning the stability of the optical frequency of the ultra-stable laser was effectively transferred to the combs. The repetition frequency of the two combs were set with a difference to produce a periodic optical interferometry through the LOS. The 16-km atmosphere link was actually a folded 8-km atmosphere link. Two telescopes  were located on the roof of our laboratory building, and a 500-mm-diameter plane mirror at a room eight kilometers away. The two combs then had to pass through several necessary fibers to the telescope, and a number of negative dispersion fibers were used to compensate for the entire dispersion of the link. Fig.~\ref{Fig:Setup} presents the experimental setup.

% Parameters
%% 249 comb is Alice, 250 comb is Bob
Both our combs contained a high-power amplifier. The repetition rate of Alice's comb was defined as 250~MHz, that is, we split the repetition frequency by a factor of 25 and used it as the 10~MHz reference clock for all our electrical systems. Meanwhile, the repetition frequency of Bob's comb was 2.6~kHz higher than that of Alice. We used a 20-nm band-pass filter centered at 1,520~nm to filter out the part with the highest intensity in the comb's spectrum. The band was lower than the Nyquist frequency of LOS to ensure the LOS did not produce any aliasing. The reflected light of the filter then passed through a 0.8-nm narrow filter centered at 1,550.12~nm and beat with the 1,550.12-nm ultra-stable laser. The linewidth of the ultra-stable laser was sub-Hz and the stability was $3\times10^{-15}$ at 1 second. In view of this optical lock, the stability of the comb's repetition rate was equal to the stability of the optical frequency of the ultra-stable laser.

To compensate for the large link loss, we had to minimize the local optic loss with some consideration of the limited laser power. The detailed structure of this module is shown in Fig.~\ref{Fig:Setup}c. First, we noted that the intensity for the local oscillator of LOS only required around 200~$\mu W$, meaning the ratio of first beam splitter (BS) after the 20-nm band-pass filter was 99:1. Almost the whole of the comb's beam went to the two-way port. Then, to split the forward signal and backward signal, a circulator was used. Compared with the original setup in Giorgetta's work~\cite{giorgetta2013optical}, which used a beam splitter to separate the local laser and the signal, our setup for LOS saved 3~dB for both forward and backward signal, thus ensuring a promotion of 6~dB for the entire link. Due to a certain amount of insertion loss with fiber components, we determined the forward loss of the box from the comb input port to the two-way port to be 2.5~dB and the backward loss of the box from the two-way port to the beam splitter for LOS to be 3~dB. This value of loss could be further suppressed if we use all free space optics to minimize the coupling loss in future work.

In our experiment, the laser power in the 20-nm band was around 120~mW and the laser power at the output port of the two-way port of the LOS module was around 70~mW. The mean value of loss for the atmosphere link itself was around 52~dB. We added an adjustable attenuator and then carefully adjusted its insertion loss. The signal power measured at the two-way port of another LOS module was found to be around 30~nW. Thus, the total loss of the link in our experiment was around 64 dB. The sensitivity of the LOS, measured using non-amplified combs locally, was found to be 3~nW. For our experiment, the spectrum of the two combs following amplification was inconsistent, and the dispersion was not as good as it was in local tests. Thus, the actual sensitivity of the LOS in our free space experiment was around 10~nW.

\begin{figure}
\centering
\includegraphics[width=0.9\columnwidth]{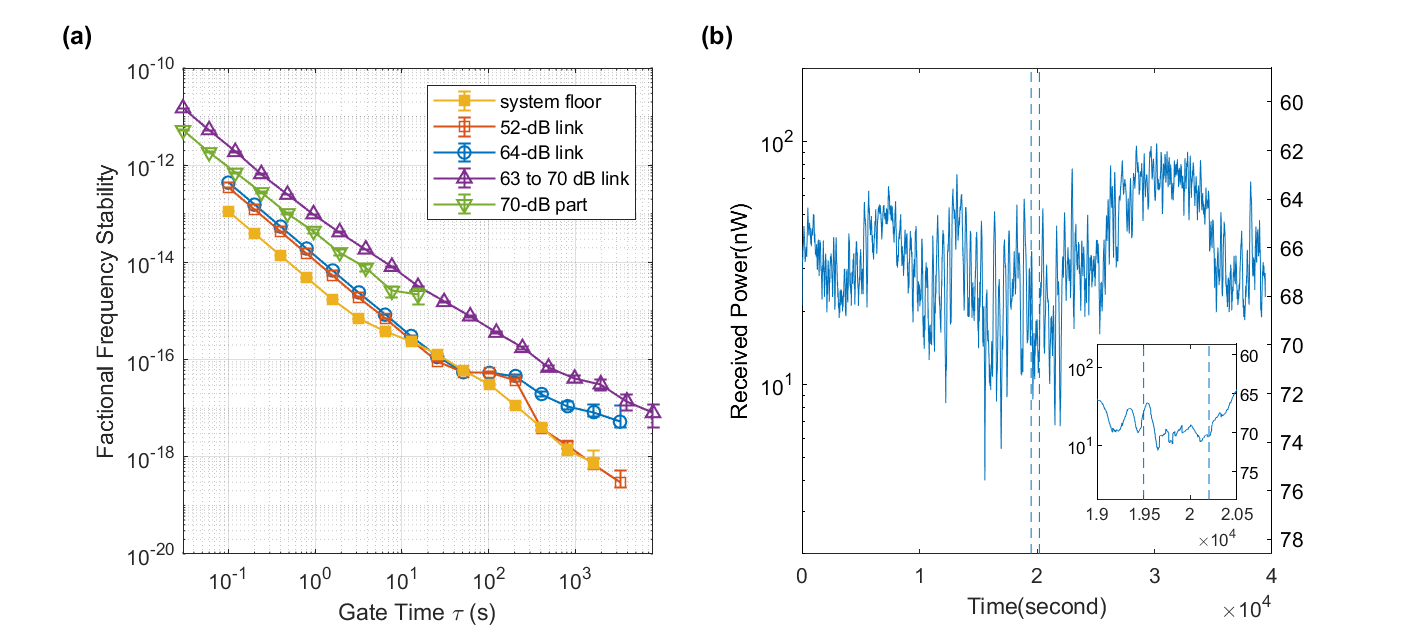}
\caption{(a). (Color online) The modified Allan deviation of frequency transfer. The line of blue open circle represents the 64-dB atmosphere link, the line of red open square represents the 52-dB atmosphere link. The local system floor is represented by the purple solid square line. The purple  triangle represents the 70-dB atmosphere link, and the green inverted triangle represents the subset data of 70-dB atmosphere link. We use an algorithm to process raw data and filter out the spikes due to sudden noise of the link induced by atmospheric scintillation. A ratio of 93\% of raw data is used to calculate the stability of the 70-dB link. This ratio is 98.3\% for the subset data in green color. (b) The received power calculated according to the power at the monitor port. The power is smoothed using ``movemean'' method and the average length is 100 seconds. The area between two dashed lines correspond to the subset data, and the average loss at this time area is 70~dB.}
\label{Fig:Result}
\end{figure}

% Results
The interferogram in our experiment updates every 380 $\mu s$, which is exactly $\frac{1}{\Delta f_r}$. Here, $\Delta f_r=2.6kHz$ is the difference between the repetition rate of the two combs. For simplicity, we used one acquisition card with two channels to record both Alice's and Bob's interferograms. Limited by the transmitted speed from the acquisition card to the computer, we only recorded 10 frames every second. Then we extracted 1024 samples from the middle of each interferogram. Further algorithms including Fast Fourier Transform (FFT), phase unwrapping, and linear fitting algorithms, provided the final arrival time. After removing certain invalid points, we obtained the modified Allan deviation (MDEV) of this frequency transfer link (Fig.~\ref{Fig:Result}a). The stability went to levels of $10^{-18}$ after 1,000 seconds. We also drew the MDEV of the 52-dB atmosphere link (without extra attenuation) and the system floor (without atmosphere link) for comparison. The frequent signal fades in the atmosphere influence the short-term stability, while the long-term stability is determined by the temperature drift of the non-common optical path. The abnormal trend near the hundreds of seconds points was a result of the periodic temperature fluctuation in our laboratory, which was caused by the air conditioning system. 

% Result of 9.13

To simulate the high loss of the uplink of the GEO satellite, the power of the comb had to be further improved. An additional comb amplifier with a two-level amplification was thus employed in our experimental setup to test the frequency transfer with a 70-dB link. The final output power was 250~mW. Following amplification, the spectrum of the comb became narrower, with the width of whole spectrum below 20~nm and the center at 1,545~nm. Thus, we replaced the band-pass filter at the start of the LOS module with a 98:2 beam splitter. The power at the two-way port of the LOS module was 145~mW. The average received power calculated from the monitered power, as shown in Fig.~\ref{Fig:Result}b, was mostly between 10~nW and 100~nW, which corresponds with 71.6~dB and 61.6~dB.  The stability reached $7.9\times10^{-18}$ when the integration time was near 8000 seconds (Fig.~\ref{Fig:Result}a). We also choose a subset of data from 19700th seconds to 20200th seconds, see dashed lines in Fig.~\ref{Fig:Result}b, where the average loss is 70~dB. The stability of these subset data is shown in green color in Fig.~\ref{Fig:Result}a. The corresponding stability is even better than that of the whole measurement, as shown in purple color in Fig.~\ref{Fig:Result}a, indicating that the link performance is not limited by the loss. We found that this is due to some bad points are included in the whole data. 

In order to tolerate higher channel loss, one direct way is to increase the optical power of the frequency comb. A high power, however, will induce large nonlinear effects, which will increase the equivalent attenuation of the link, and reduce the performance of the LOS. It would be useful to study how to minimize the nonlinear effect at even higher powers. Meanwhile, to further reduce the instability, we packed the non-common optical path into a small box, and stabilized its temperature using thermo-electric coolers (TEC). A number of further optimizations need to be carried out, including flatterning the spectrum of the comb to improve the measurement precision, which will, in turn, help to improve the performance with feedback. 
 
In summary, we demonstrated the optical frequency transfer through an up to 70-dB free space channel with an instability of orders of $10^{-18}$. This loss was equivalent to the loss of the two-way satellite-ground links for the GEO satellite. Moreover, the technology developed in our experiment can also find immediate applications in quantum communication, like free space twin-field quantum key distribution~\cite{chen2020sending} and quantum repeater~\cite{duan2001long}. 

%In summary, we demonstrated the optical frequency transfer through an up to 70-dB free space channel with an instability of orders of $10^{-18}$. This loss was equivalent to the loss of the two-way satellite-ground links for the GEO satellite. While higher optical power can resist more channel loss, a high power will induce large nonlinear effects, which will increase the equivalent attenuation of the link, and reduce the performance of the LOS. It would prove useful to study how to minimize the nonlinear effect at even higher powers. Meanwhile, to further reduce the instability, we packed the non-common optical path into a small box, and stabilized its temperature using thermo-electric coolers (TEC). A number of further optimizations need to be carried out, including flatterning the spectrum of the comb to improve the measurement precision, which will, in turn, help to improve the performance with feedback. Future steps towards the satellite-ground time-frequency transfer should include incorporating the Doppler effect with the high loss link and testing all instruments through various space examinations.

\section{Methods}

\textbf{Satellite ground link loss analysis.} The link loss was quite different for the downlink ($\eta_{down}$) and the uplink ($\eta_{up}$), which represent satellite-to-ground and ground-to-satellite, respectively, as was shown in Fig.~\ref{Fig:height} and Table 1. Thess losses can be calculated using Eq.~\ref{link loss},

\begin{equation}\label{link loss}
\begin{split}
\eta_{down}=\eta_{tele\_s}(\frac{D_{g}}{L \theta_{down}})^2 T_{atm} \eta_{tele\_g} \eta_{sm\_g},\\
\eta_{up}=\eta_{tele\_g} (\frac{D_{s}}{L \theta_{up}})^2 T_{atm} \eta_{tele\_s}\eta_{sm\_s} ,
\end{split}
\end{equation}

where $D_{g}$ and $D_{s}$ are the telescope aperatures of ground and satellite, respectively, L is the satellite-to-ground distance. Supposing that the diameter for both telescopes is approximately 1 m, a typical dimension for optical space telescopes, $\theta_{down}$ and $\theta_{up}$ represent the effective transmitter full-angle divergence for the downlink and the uplink, respectively. It should be noted that the divergence angle not only includes the transmitter telescope divergence, but also the influence of atmospheric turbulence and the pointing error of the acquiring, pointing and tracking (APT) system. In transmission, the optical beam will be broadened or deflected by diffraction, air turbulence and mispointing, which are referred to here as diffraction loss. The diffraction loss of the uplink was $\left(\frac{D_{s}}{L\theta_{up}}\right)^2$, while for the downlink, this is $\left(\frac{D_{g}}{L\theta_{down}}\right)^2$. In the uplink, atmospheric turbulence close to the ground strongly deteriorates the quality of the beam. Therefore, turbulence-induced distortion will significantly increase the beam divergence angle. The divergence angle is about 22~$\mu rad$ according to our previous result related to the `Micius` satellite~\cite{ren2017ground}. The atmospheric seeing in observatory stations is commonly smaller than 10 urad. With the adaptive optics system, the effects of turbulence can be partially compensated for \cite{Vasylyev2016}. The pointing error of the APT system will lead to spot position jitter, which also influences the effective divergence angle. This effect can be smaller for satellite in higher orbits due to the smaller angular velocity and acceleration than with satellites in low orbits. Thus, it is reasonable to obtain ~15 urad of $\theta_{up}$ with a comparatively large transmitter telescope aperture and optimized adaptive optics. In the downlink, the beam reaches the air turbulence with a large size and is received immediately after crossing the atmosphere, thus, the impact of air is much smaller for the beam broadening compared to that of the uplink, meaning the effective divergence angle can be close to the diffraction limit. With the optical wavelength of 1550 nm and the diameter of telescope on satellite is 1 m, the divergence angel of  4 urad for $\theta_{down}$ is technically achievable \cite{Cao2018}.

The telescope optical efficiency of satellite and ground are $\eta_{tele\_s}$ and  $\eta_{tele\_g}$, respectively, both of which were set to 0.8. Given the two-way link symmetry, the satellite telescope was the transmitter for the downlink and the reciever for the uplink.  $T_{atm}$ is the atmosphere transmittance, which is reduced by the air absorption and the scattering of the propagating beam. The attendant value is related to the visibility, the altitude and the optical wavelength, etc., with the typical value around 0.7 for 1,550 nm at a sea-level location with 5-km visibility\cite{Bourgoin2013}. Meanwhile, $\eta_{sm\_s}$ and $\eta_{sm\_g}$ represent the single-mode fiber coupling efficiency on the satellite and on the ground, respectively. In terms of the ground, for the downlink, the adaptive optics technology could be employed to optimize the coupling efficiency to over 15$\%$ \cite{jovanovic2016efficiently}, while for the uplink, since the receiver was in a high-speed moving satellite and the single-mode fiber receiving field of view was very small, the single-mode coupling efficiency was very sensitive to the optical axis consistency between the ground and the satellite telescope. This was estimated to be 0.05 for uplink. The comparison of the links for each type of satellite is shown in Table 1. 

\textbf{Frequency locking.} In our setup, the reference frequency was determined via the stable laser at Alice's side. We can write the relationship of frequency as $f_{laser,A}=f_{CEO,A}+M*f_{rep,A}+f_{beat,A}$, where $f_{laser,A}$ is the frequency of the stable laser, $f_{CEO,A}$ is the CEO frequency, $f_{rep,A}$ is the repetition frequency, and the $f_{beat,A}$ the beat frequency between the comb teeth and the stable laser. Because the stable laser was the only clock in our system, the repetition rate of the comb should only be determined according to the frequency of the stable laser. That is, a small drift of carrier-envelope offset (CEO) frequency must not influence the repetition rate. To achieve this goal, the ideal method involves using self-referencing, which lock the both $f_{CEO,A}$ and $f_{beat,A}$ to a frequency proportion to $f_{rep,A}$. Here, we used an easier method that required no modification in terms of the default configuration of the frequency comb. The frequency comb uses an internal rubidium clock as a reference clock for all its electrics. We set the CEO frequency and the beat frequency to the same value and adjusted their sign to make $f_{CEO,A}=-f_{beat,A}$. Thus, we had $f_{laser,A}=M*f_{rep,A}$, which means the repetition frequency of the comb was only related to the optical frequency of the stable laser. The same approach was then used on Bob's side.

After locking the comb to the stable lasers, the two lasers at each side had to be locked such that they represented the same clock. Here, the "same clock" does not mean that the difference between the absolute frequency of the two lasers was always fixed; rather, it means that the frequency of the second laser was always proportional to the frequency of the first. When Alice's stable laser had a drift of 10~kHz, Bob's stable laser had to have a corresponding shift, albeit not exactly 10~kHz. The simplest solution here was to ensure the local oscillator synchronized with the repetition rate of Alice's comb. Because $f_{laser,A}=M*f_{rep,A}$, we had $f_{laser,B}-f_{laser,A}\propto f_{rep,A}\propto f_{laser,A}$, meaning the two stable lasers could be regarded as synchronized.

\textbf{LOS and time offset extraction}. One way to extract the high-precision time is through LOS, which required that the two optical combs had a slight difference in terms of repetition rate. When they beat with each other, the pulse from one comb slowly stepped by another, generating a series of inteference points with difference voltages. We refer to these points as a `frame`. The interference signals then had to go through a low-pass filter and could thus be recorded for further analysis. To clearly show how to extract the time offset, it is useful to analyze it within the frequency domain. We can write the electric field of the two combs as follows:
\begin{equation}
E_A(t)=\exp(i2\pi \omega_A t)\sum_n\left[E_{A,n}\exp(i2\pi nf_r t)\right],
\end{equation}
\begin{equation}
E_B(t)=\exp(i2\pi \omega_B t)\sum_n\left[E_{B,n}\exp\left(i2\pi n(f_r+\Delta f_r)t\right)\right].
\end{equation}
Here, $\omega_A$ and $\omega_B$ are the center frequency of the combs in Alice's and Bob's sides. The repetition frequency of Alice's comb was $f_r$, while Bob's comb had a slightly higher repetition frequency of $f_r+\Delta f_r$. After the low-pass filter with bandwidth $f_r/2$, only the beat of most of the adjacent frequencies was maintained. Thus the voltage was as follows:
\begin{equation}\label{inf:0}
\begin{split}
V(t)\propto & \mbox{Im}\left[E_A(t)^*E_B(t)\right]=\exp[i2\pi(\omega_B-\omega_A)t]\\
&\times \sum_n E_{A,n}E_{B,n}\exp[2i\pi n\Delta f_r t].
\end{split}
\end{equation}
Now, supposing that Bob's comb had a delay $\tau$, note that $\tau\in[-\frac{1}{2f_r},\frac{1}{2f_r})$, otherwise, we could align the pulse of Alice's comb to another pulse of Bob's. This delay resulted in the interferogram experiencing a temporal shift, and the corresponding phase of the frequency spectrum underwent a change in slope. We can write the new interferogram as follows:
\begin{equation}\label{inf:tau}
\begin{split}
V(t,\tau)\propto & \mbox{Im}\left[E_A(t)^*E_B(t-\tau)\right]=\exp\{i2\pi[(\omega_B-\omega_A)t-\omega_B\tau]\}\\
&\times \sum_n E_{A,n}E_{B,n}\exp[2i\pi(n\Delta f_r t-n(f_r+\Delta f_r)\tau)].
\end{split}
\end{equation}
We use $S(\nu,\tau)$ to represent the Fourier transform of the interferogram, while $\angle S(\nu,\tau)$ stands for the phase of the frequency spectrum. From equations \ref{inf:0} and \ref{inf:tau}, we have,
\begin{equation}\label{LOS}
\begin{split}
\angle S(\nu,0)&=0\\
\angle S(\nu,\tau)&=-2\pi\tau\frac{f_r+\Delta f_r}{\Delta f_r}[\nu-(\omega_B-\omega_A)]-2\pi\omega_B\tau.
\end{split}
\end{equation}
The phase is a linear function to $\nu$, and the slope of frequency spectrum $-2\pi\tau\frac{f_r}{\Delta f_r}$ is proportional to the delay $\tau$.

After the calculation of the precise delay $\tau$, the clock offset $\Delta t$ can be easily derived through the principle of two-way time synchronization: $T_{A\rightarrow B}=T_{Link}-\Delta t$, $T_{B\rightarrow A}=T_{Link}+\Delta t$. The atmosphere usually suffers low-frequency noise such as wind and turbulence, meaning $T_{Link}$ is stable during a ten-millisecond period. Thus, we have,
\begin{equation}
\Delta t=\frac{1}{2}(T_{B\rightarrow A}-T_{A\rightarrow B}).
\end{equation}

The linear optical sampling requires a fixed frequency difference for the two combs. From Eq.~\ref{LOS}, it is clear that the slope is inversely proportional to the $\Delta f_r$. Given a specific interferogram, the delay time $\tau$ is proportional to the pre-set value of $\Delta f_r$. An incorrect value of $\Delta f_r$ can lead to a continuous drift of $\tau$, and this will soon exceed the ambiguous range of $\frac{1}{f_r}$. Thus, a real frequency transfer using LOS must be used to calibrate the frequency difference. Here, because our two stable lasers were locked with each other, we could obtain the frequency difference through a simple calculation. At first, we have $f_{laser,A}=M*f_{rep,A}$ and $f_{laser,B}=N*f_{rep,B}$. Using a wavelength meter that is accurate to 10~MHz, the value of $M$ and $N$ can be precisely determined. Because the $f_{rep,A}=250~MHz$, the $f_{laser,A}$ can be calculated. Then, we get the value of $f_{rep,B}$ and $f_{laser,B}$. The parameters are listed in Table 1.

\begin{table}\label{Table:Freq}
\centering
\begin{tabular}{|c|c|}
\hline
$f_{rep,A}$ & 250~MHz \\ \hline
M			& 773,603 \\ \hline
$f_{laser,A}$ & 193.400,75~THz \\ \hline
$f_{rep,B}$ & 250.002,600,8~MHz \\ \hline
N 			& 773,592        \\ \hline
$f_{laser,B}$ & 193.400,011,96~THz \\ \hline
$\Delta f_r$ & 2,600.802,5~Hz \\ \hline
\end{tabular}
\caption{The parameters of our experiment. The frequency reference is the repetition rate of Alice's comb, which means $f_{rep,A}$ is defined as 250~MHz.}
\end{table}

After calculating the $\Delta f_r$, we had to extract the time offset and track its change. For our frequency transfer experiment, we ignored the initial time offset of the two clocks, only focusing on the relative change in time offset. Thus, for each frame, we only extracted the time offset relative to the first frame, which involved two processes. During the first process, we carried out FFT analysis on the frame data to obtain the phase information. Then we calculated the phase difference between this frame and the first frame. The slope and the time offset $\tau$ could then be determined. In this step, the influence of system dispersion was also canceled out. For the second process, we calculated the time offset value for an ideal clock at that specific frame. More specifically, because the interval of the adjacent pulse for the two combs was different, for every pulse of Alice's comb, the corresponding pulse of Bob's comb would go a bit quicker. The difference was $dt=\frac{\Delta f_r}{f_r(f_r+\Delta f_r)}$. Given the start position for the first frame and this frame, named $p_{1}$ and $p_{N}$, in the unit of the interval of Alice's comb, we know that the theoretical time offset provided by Bob's comb was always accurate. The value was $\tau_{theo}=\mbox{mod}((p_N-p_1)\times dt, \frac{1}{f_r})$. Thus, we had to subtract this from the value of $\tau$. In other words, the value of $\tau=\tau_{theo}$ as calculated in the first process, actually means the time offset was not changed.

Finally, it is important to pay some attention to the sign of the time offset. We can recall that only one acquisition card was used to record two LOS signals. The LOS at Alice's side meant the arrived time of Bob's signal had a delay equal to $T_{Link}$, which gives a negative slope, while at Bob's side, the arrived time of Alice's signal had a delay equal to $T_{Link}$, which was equivalent to that of the arrived time of Bob's signal $T_{Link}$ ahead. The LOS at Bob's side now finally presents a positive slope. Thus, we should flip the sign of $tau$ at Bob's side. Finally, the formula $\Delta t=\frac{1}{2}(T_{B\rightarrow A}-T_{A\rightarrow B})$ provided the correct change in time offset during the whole experiment.

%\section{Data availability}
%The data that support the plots within this paper and other findings of this study are available from the corresponding author upon reasonable request.
\section*{Acknowledgements}
The authors thank enlightening discussions with Liang Zhang and Xuan Zhang. This work was supported by the Strategic Priority Research Program on Space Science of the Chinese Academy of Sciences, the National Key R\&D Program of China (2017YFA0303900), the National Natural Science Foundation of China, the Anhui Initiative in Quantum Information Technologies (AHY010100), the Key R\&D Program of Guangdong Province (2018B030325001), and the Special Development Fund of Zhangjiang Natiional Innovation Demonstration Zone.

%\section*{References}

\bibliographystyle{apsrev4-1}

\bibliography{FT_Reference}

%\section*{Authors' contributions}
%All authors contributed extensively to the work presented in this paper.
%\section*{Competing interests}
%The authors declare no competing interests.

%\section{Authors' information}

\end{document}